\documentclass[onecolumn]{emulateapj}
\usepackage{natbib}

\shorttitle{M31 pixel lensing PLAN campaign}
\shortauthors{Calchi Novati et al.}

\begin{document}

\title{M31 pixel lensing PLAN campaign: MACHO lensing and Self lensing signals}

\author{
S.~Calchi Novati\altaffilmark{1,2}\footnote{mailto: novati@sa.infn.it},
V.~Bozza\altaffilmark{2,3},
I.~Bruni\altaffilmark{4},
M.~Dall'Ora\altaffilmark{5},
F.~De Paolis\altaffilmark{6,7},
M.~Dominik\altaffilmark{8}\footnote{Royal Society University Research Fellow},
R.~Gualandi\altaffilmark{4},
G.~Ingrosso\altaffilmark{6,7}, 
Ph.~Jetzer\altaffilmark{9},
L.~Mancini\altaffilmark{10,2}, 
A.~Nucita\altaffilmark{6,7},
M.~Safonova\altaffilmark{11},
G.~Scarpetta\altaffilmark{1,2,3},
M.~Sereno\altaffilmark{12,13}, 
F.~Strafella\altaffilmark{6,7},
A.~Subramaniam\altaffilmark{11},
A.~Gould\altaffilmark{14}\\
(PLAN collaboration)}

\affil{$^1$ Istituto Internazionale 
per gli Alti Studi Scientifici (IIASS), 
Via Pellegrino 19, 84019 Vietri Sul Mare (SA), Italy}
\affil{$^2$ Dipartimento di Fisica ``E. R. Caianiello'', 
Universit\`a di Salerno, Via Giovanni Paolo II 132, 84084 Fisciano (SA), Italy}
\affil{$^3$ INFN, Sezione di Napoli, Via Cinthia 9, 80126 Napoli, Italy}
\affil{$^4$ INAF, Osservatorio Astronomico di Bologna, 
Via Ranzani 1, 40127 Bologna, Italy}
\affil{$^5$ INAF, Osservatorio Astronomico di Capodimonte, 
Salita Moiariello 16, 80131 Napoli, Italy}
\affil{$^6$ Dipartimento di Matematica e  Fisica ``E. De Giorgi'', 
Universit\`a del Salento, CP 193, 73100 Lecce, Italy}
\affil{$^7$ INFN, Sezione di Lecce, Via Arnesano, 73100 Lecce, Italy}
\affil{$^8$ SUPA, University of St Andrews, School of Physics \& Astronomy, North
Haugh, St Andrews, KY16 9SS, United Kingdom}
\affil{$^9$ Institute for Theoretical Physics,  University of Z\"urich, 
Winterthurerstrasse 190, 8057 Z\"urich, Switzerland}
\affil{$^{10}$ Max Planck Institute for Astronomy, K\"onigstuhl 17, 
69117 Heidelberg, Germany}
\affil{$^{11}$ Indian Institute of Astrophysics, Bangalore 560 034, India}
\affil{$^{12}$ Dipartimento di Scienza Applicata e Tecnologia, 
Politecnico di Torino,  Corso Duca degli Abruzzi 24, 10129 Torino, Italy}
\affil{$^{13}$ INFN, Sezione di Torino, 
Via Pietro Giuria 1, 10125, Torino, Italy}
\affil{$^{14}$ Department of Astronomy, 
Ohio State University, 140 West 18th Avenue, Columbus, OH 43210, US}

\begin{abstract}
We present the final analysis of the observational campaign carried 
out by the PLAN (Pixel Lensing Andromeda) collaboration to detect 
a dark matter signal in form of MACHOs through the microlensing effect.
The campaign  consists  of about 1~month/year observations carried out during 4 years
(2007-2010) at the 1.5m Cassini telescope in Loiano 
(''Astronomical Observatory of BOLOGNA'', OAB) plus 
10~days of data taken in 2010 at the 2m Himalayan Chandra Telescope (HCT)
monitoring the central part of M31 (two fields of about $13'\times 12.6'$). 
We establish a fully automated pipeline for the search
and the characterization of microlensing flux variations: as a result
we detect 3~microlensing candidates. We evaluate the expected
signal through a full Monte Carlo simulation of the experiment
completed by an analysis of the detection efficiency of our pipeline.
We consider both ``self lensing'' and ``MACHO lensing'' lens populations,
given by M31 stars and  dark matter halo MACHOs, in the M31 and the Milky Way (MW), respectively.
The total number of events is compatible with the expected self-lensing rate.
Specifically,  we evaluate an expected signal of about 2 self-lensing events.
As for MACHO lensing, for full $0.5~(10^{-2})~\mathrm{M}_\odot$ MACHO halos, 
our prediction is for about 4 (7) events. 
The comparatively small number of expected
MACHO versus self lensing events, together with the small number statistics
at disposal, do not enable us to put strong constraints
on that population. Rather,  the hypothesis, suggested by a previous analysis,
on the MACHO nature of OAB-07-N2, one of the microlensing candidates,
translates into a sizeable \emph{lower} limit for the halo mass
fraction in form of the would be MACHO population, $f$, of about 15\%
for $0.5~\mathrm{M}_\odot$ MACHOs. 
\end{abstract}


\section{Introduction} 
Gravitational microlensing \citep{roulet97} is the tool of choice
for the investigation of the dark matter content 
of galactic halos \citep{strigari13} in form of compact objects (MACHOs).
Since the original paper of \cite{pacz86},
observational campaigns have been undertaken to this purpose
towards the Magellanic Clouds \citep{moniez10}, as a probe of the MW halo, 
and towards the nearby galaxy of Andromeda, M31 \citep{grg10}.
Although there is an agreement in excluding that
MACHOs can fill up the dark matter halos,
some tension remains based on the difficulty to fully disentangle
the lensing signal from known (stellar) population
(``self lensing'') as opposed to
the dark matter signal (MACHO lensing).

The EROS \citep{eros07} and more recently the
OGLE collaboration \citep{ogle09,ogle10,ogle11,ogle11b},
out of observations towards the Large and Small
Magellanic Clouds (LMC and SMC),
put rather robust upper limits (at 95\% CL) 
on the halo mass fraction in form of MACHO, $f$,
below $10\%$ up to $1~\mathrm{M}_\odot$ MACHOs
(and down to below $5\%$ around $10^{-2}~\mathrm{M}_\odot$ objects).
On the other hand, the MACHO collaboration had
reported a MACHO signal towards the LMC of about $f\sim 20\%$ within
the mass range $(0.1-1)~\mathrm{M}_\odot$ \citep{macho00,bennett05}.

To address the reasons of this disagreement
the self-lensing scenario, originally discussed in \cite{sahu94,wu94,gould95},
has been thereafter the object of several analyses \citep{derujula95,aubourg99,
salati99,alves00,gyuk00,jetzer02,
mancini04,griest05,novati06,novati09b,novati11,novati13}. Alternative hypotheses
have also been discussed, in particular proposing
non-standard models of the LMC/SMC which may somehow
enhance the expected self-lensing rate \citep{zaritsky97,zhao98,gould98,evans00,zaritsky04,besla13}. 

The main bonus of the line of sight towards M31 \citep{crotts92,agape93,jetzer94} is that,
being an external galaxy, we can fully map its own dark matter halo
(roughly, at parity of MACHO mass function and halo fraction,
one expects about 2/3 of the MACHO signal, if any, 
to belong to the M31 halo, with the rest to the MW halo along that line of sight).
Because of the large ($770~\mathrm{kpc}$) distance to the sources
we enter here the ``pixel lensing'' regime of microlensing \citep{gould96}.
In particular, among other specificities, we 
recall the further degeneracy in the lensing
parameter space between the physical event duration,
the Einstein time, $t_\mathrm{E}$ and the impact parameter, $u_0$, which makes
reliable, in most cases, only a determination
of the full-width-half-maximum duration, $t_\mathrm{FWHM}$,
\citep{gould96,wozniak97,gondolo99,alard01,riffeser06,dominik09}.
Additionally, as further addressed below, it results that
the ratio of the expected self lensing over
MACHO lensing rate is larger with respect to that expected towards the LMC/SMC
(quantitatively this depends on the field of view and on the assumed 
MACHO mass function)
and this further complicates the physical interpretation
of the data along this line of sight.
Indeed, the analysis of the self-lensing signal
appears to be at the origin of the disagreement between
the POINT-AGAPE collaboration \citep{novati05},
who reported an evidence for a MACHO signal
(a different analysis of POINT-AGAPE is discussed in \citealt{belokurov05}),
and the MEGA collaboration \citep{mega06}
(but see also \citealt{mega04})
who concluded that their signal
could be fully explained by the expected self-lensing rate
(see also the further analyses in \citealt{ingrosso06,ingrosso07}).

Following the difficulty to disentangle the MACHO and the self-lensing signals
by considering full sets of events, the detailed
analysis of single events turn out to be very important.
Interestingly all the analyses of this kind presented up to now, concerning three distinct
microlensing candidates towards M31, indicate that the lens should
more likely be attributed to the MACHO lensing population
\citep{point01,riffeser08,novati10}. 

The more recently undertaken M31 pixel lensing PAndromeda survey
\citep{lee12}, which by large overtakes previous ones in term
of monitored field of view, baseline extension and cadence
(all essential issues to both enhance the expected rate and
well characterize the signal) and out of which
the detection of 6 new microlensing candidates
out of a first analysis of their first year of observation
has been reported, promises to mark an important step forward in this framework.

As PLAN  collaboration we have undertaken
a pixel lensing survey campaign towards M31
based at the Cassini telescope in Loiano (OAB).
Following a first pilot season with 11 consecutive nights of observations
in 2006 \citep{novati07}, which essentialy probed
the feasibility of the project, we have then undertaken
a campaign eventually lasted four years, 2007-2010.
In 2010 we have extended the monitoring
to the 2m Himalayan Chandra Telescope (HCT).
The results of the 2007 campaign have been discussed in \cite{novati09},
with the presentation of a fully automated selection pipeline
out of which we had selected 2 new microlensing candidates,
that we dubbed OAB-N1 and OAB-N2
which we are now going to refer to as OAB-07-N1 and OAB-07-N2, with the additional indication
of the year of detection),
with OAB-07-N2 being then the object
of a further analysis, including that
of the lens proper motion
(also thanks to additional data kindly provided by the WeCAPP collaboration,
\citealt{wecapp01,wecapp03}), presented in \cite{novati10}.

In the present  work we intend to present the final analysis
of the PLAN survey including all four years of observations,
both OAB and HCT data. In particular, we present
a third microlensing candidate, already presented
in \cite{lee12}, discuss the expected signal,
both self lensing and MACHO lensing,
and compare it to the observed rate.
In Sect.~\ref{sec:setup} we present the observational data;
in Sect.~\ref{sec:data} we highlight the main steps
of data reduction and our photometry procedure; 
in Sect.~\ref{sec:pipe} we outline the
method of our automated pipeline
and present the results for the search
of microlensing candidates;
in Sect.~\ref{sec:mls} we discuss
our analysis to establish the expected signal:
a full Monte Carlo simulation of the experiment
completed by an analysis of the detection efficiency
of our pipeline; in Sect.~\ref{sec:machos}
we present the expected signal and discuss the MACHO lensing
versus self-lensing issue as compared to the observed rate;
in Sect.~\ref{sec:end} we present our conclusions.

\section{M31 PLAN pixel lensing survey} \label{sec:ana}

\subsection{Observational data} \label{sec:setup}

\begin{figure}
\epsscale{1.}
\plotone{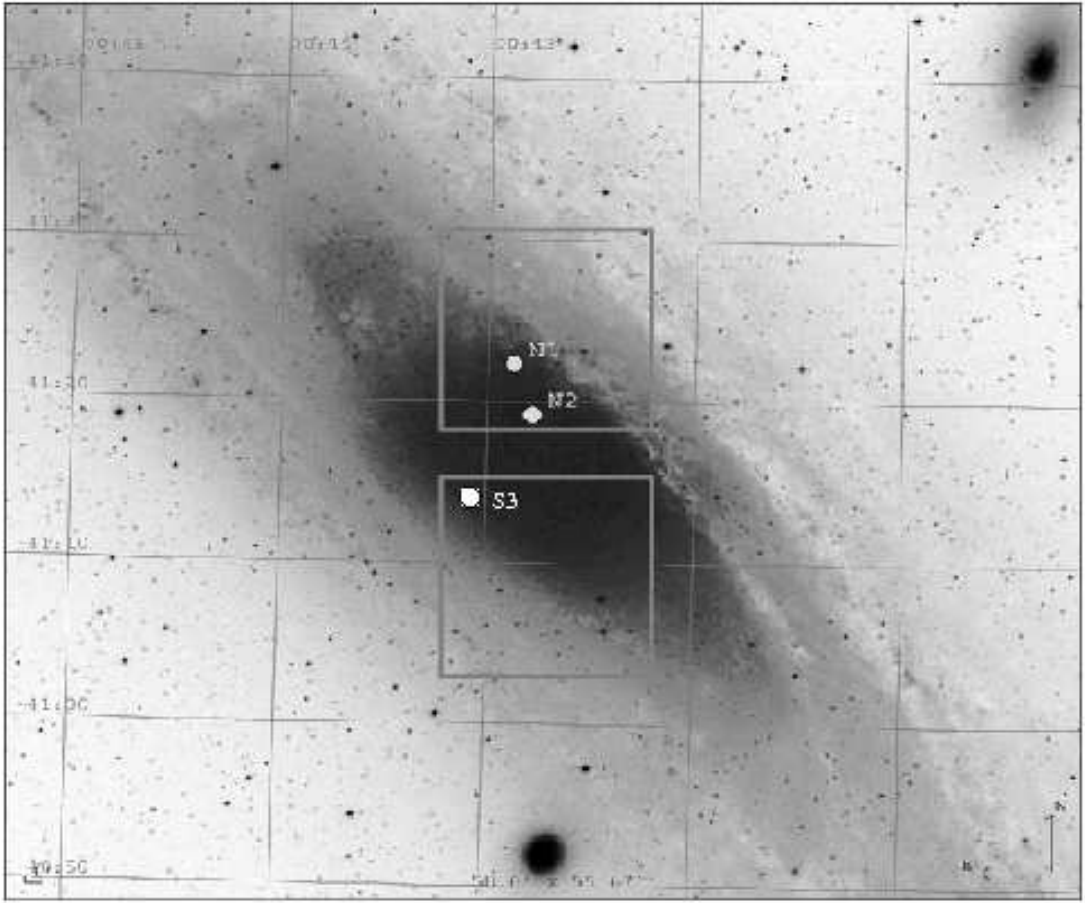}
\caption{
Projected on a background archive image of M31
we display the fields of view of the OAB 2007-2010 PLAN pixel
lensing campaign and the positions of our 3 microlensing
candidates: OAB-07-N1, OAB-07-N2 \citep{novati09} and OAB-10-S3,
first appeared in \cite{lee12} and known as PAnd-4.
\label{fig:m31}
}
\end{figure}

Most data of our pixel lensing campaign have been collected
at the 1.5m Cassini telescope in Loiano, ``Osservatorio
Astronomico di Bologna'' (OAB, http://www.bo.astro.it/loiano/),
785~m above sea level nearby the city of Bologna (Italy).
The photometric monitoring was carried out
using the ``Bologna Faint Object Spectrograph and Camera'' (BFOSC)
equipped with a CCD EEV LN/1300-EB/1 back illuminated and AR coated,
read out noise 3.1~e$^-$/pixel and gain 2.22 ADU/pixel,
pixel scale of $0.58''$/pixel and with 
$1340 \times 1300$ pixels  for
a total field of view of $13'\times 12.6'$.
We monitored two fields of view around the inner M31 region,
centered at $\mathrm{R.A}=0^\mathrm{h} 42^\mathrm{m} 50^\mathrm{s}$,
and $\mathrm{decl.}= 41^\circ 23' 57''$ (North) and  $\mathrm{decl.}= 41^\circ 08' 23''$
(South), with axes parallel to the south-north and
east-west directions, just leaving out the very inner M31 bulge region (Fig.~\ref{fig:m31}).
The data have been collected in two broad band filters
(similar to Cousin $R$ and $I$), with exposure times
up to 10~minutes per frame depending on the filter
and on the moon level. The standard data reduction  
have been carried out within the IRAF package (http://iraf.noao.edu/). 

\begin{center}
\begin{deluxetable}{ccccc}
\tablecolumns{4}
\tablewidth{0pc}
\tablecaption{Observational sampling for the 2006-2010 OAB campaign.
The first year pilot season is then no longer considered within the analysis.
\label{tab:sampling}}
\tablehead{
\colhead{(1)} & \colhead{(2)} & \colhead{(3)} & \colhead{(4)} & \colhead{(5)}\\
}
\startdata
2006& Sep01-Sep11 & 11 & 8 & 4.2\\
\hline
2007& Nov12-Dec31 & 50\tablenotemark{a} & 31 & 3.8\\
2008& Sep15-Nov23 & 65\tablenotemark{b} & 38 & 4.6\\
2009& Sep17-Oct22 & 36 & 25 & 5.5\\
2010& Sep20-Oct31 & 41\tablenotemark{c} & 20 & 4.6\\
\hline
tot& & 192 & 114 & 4.6\\
\enddata
\tablecomments{
(1): Year; (2): period of the year; (3): number of 
nights awarded for our project; 
(4): number of  nights with at least some M31 observations;
(5): average number of hours/night of M31 observations.
In the calculation of the last row the data from the 2006 pilot season are not included.
The PI for the 2006 and 2007 proposals was F.~Strafella,
for the 2008-2010 S.~Calchi~Novati.
In 2010 we have also taken data at the 2m HCT telescope
for 10 consecutive days, Oct01-Oct10, PI A.~Subramaniam.
}
{
\tablenotetext{a}{12 consecutive nights have been partly shared with another observer.}
\tablenotetext{b}{9 not-consecutive nights have been partly shared with other observers
and 5 full not-consecutive nights have been allocated to other observers.}
\tablenotetext{c}{1 full night have been allocated to another observer.
}
}
\end{deluxetable}
\end{center}

The typical microlensing events we expect to detect
are relatively faint flux variations (with flux deviation at maximum magnification fainter
than about $R\sim 20$) lasting up to a few days.
Given the available experimental setup, these features
fix our observational requirements. In particular
we need a long enough overall baseline with a suitable sampling
and high S/N data, namely we ask for full and consecutive nights of observations
for an overall  period up to about 2 months. 
The details of the sampling of the first pilot
season (2006), whose data are however not further considered
in the following, and for the following 
four years campaign (2007-2010) are reported in Table~\ref{tab:sampling}.
Overall, the average weather conditions (humidity, cloud coverage)
did not turn out to be optimal to our purposes, with our sampling full of 
unwanted gaps (in particular, the consequence
of the non-optimal sampling will be made apparent
by the following discussion on the failed microlensing
candidate in 2008 and the analysis of the 2010 data below). 
Indeed, the fraction of at least partially
clear nights has remained around 60\%,
with overall 114 at least partially ``good'' nights over the 192 allocated ones. 
Considering however the number of hours we could actually spend observing M31,
with an average number of visibility hours 
given the period of the year and the declination
of the site ($44^\circ.4$ North, almost ideal for observations towards M31)
up to almost 10 hours/night, the overall fraction of hours we could monitor M31
with respect to the allocated ones drops to below 40\%. 
Although the quality of the data turned out to be good enough,
still we had to reject a sizeable fraction of ``bad'' images
(very poor seeing conditions and/or too high moon level).
Specifically, within the selection pipeline we do mask data points with large relative
error bars: this further reduced the number of 
available data points to about 80-90 and 70-80,
depending on the light curves, for $R$ and $I$-band data respectively.
This number must be compared to the initial number
of awarded nights, 192, for a fraction below 50\%.
The sky brightness, because of anthropic pollution of the nearby towns, is about
$1~\mathrm{mag}$ brighter than in a typical dark site.
The typical  seeing values were around $2''$ with
a strong scatter, though, to further complicate the analysis. 

In 2010 we submitted a proposal to carry out
parallel observations to those at OAB at the 
2m Himalayan Chandra Telescope (HCT)
at the Indian Astronomical Observatory (IAO, http://www.iiap.res.in/iao/cycle.html)
at  $32^\circ$ North and located at 4500~m above sea level, PI A.~Subramaniam, and we were awarded
with 10 consecutive nights, October 1-10
(therefore, within the shift allocated at OAB)  for 2 hours/night. 
The photometric monitoring was carried out
using the ``Himalayan Faint Object Spectrograph and Camera'' (HFOSC) 
equipped with a Thompson CCD with
read out noise 4.8~e$^-$/pixel, gain 1.22 ADU/pixel,
pixel scale of $0.296''$/pixel and with 
$2048 \times 2048$ pixels  for
a total field of view of about  $10'\times 10'$,
slightly smaller than that at OAB. To match
with OAB observations we have observed
two fields, North and South the M31 center,
centered in $\mathrm{R.A}=0^\mathrm{h} 42^\mathrm{m} 50^\mathrm{s}$
and $\mathrm{decl.}= 41^\circ 22' 57''$ and $\mathrm{decl.}= 41^\circ 09' 23''$,
respectively, so to be fully included within the OAB fields of view,
with observations evenly distributed in two broad
band filters, $R_\mathrm{C}$ and $I$.
We obtained useful data out of all the 10 scheduled nights, 
with only some problems of guiding that forced
us to reduce the exposure time down to 5~minutes/exposure
against the programmed one of 10~minutes/exposure.

\subsection{Data analysis} \label{sec:data}

The raw data are first reduced with bias and (sky) flat field frames 
(plus defringing for $I$ band data)
using the \texttt{ccdred} tasks within IRAF. The photometry
is carried out according to the ``superpixel'' scheme,
first introduced within the AGAPE group \citep{agape97}
and further discussed in \cite{novati02,novati09},
which is a fixed-size aperture photometry
(we use $5\times 5$ superpixels) with a linear empirical
correction to account for the seeing variations.
Several images (up to about 20) are taken
each night per band and per field. 
The superpixel light curve
is built with a weighted (by the inverse of the square
of the flux error) average carried out  after the seeing correction
so to end up with 1 data point per night per filter.
This procedure is suitable to match the expected
typical event duration of about a few days.

For the analysis of HCT data, right after the standard CCD reduction, 
we resample them so to match the OAB pixel scale.
To this purpose, first we draw a list of about 300 reference stars per field
we use to establish the relative astrometry
and then,  using the \texttt{immatch} tasks within IRAF, we carry out the pixel resampling
(moving from the HCT $0.30''$ to the OAB $0.58''$ pixel scale).
The rms of the relative astrometry on the resampled
HCT images versus the OAB ones is at most
of 0.3 pixel. The resampled HCT data
are then processed exactly as the OAB data.

\subsection{Pixel lensing pipeline} \label{sec:pipe}

The purpose of the pipeline is to establish a list of bona fide microlensing \emph{candidates}.
Within this scheme, our specific aim is to build a \emph{fully automated} pipeline.
This is crucial to deal with large data set,
however the key aspect is that this enables
us to reliably estimate the detection efficiency.

\begin{figure}
\epsscale{1.}
\plotone{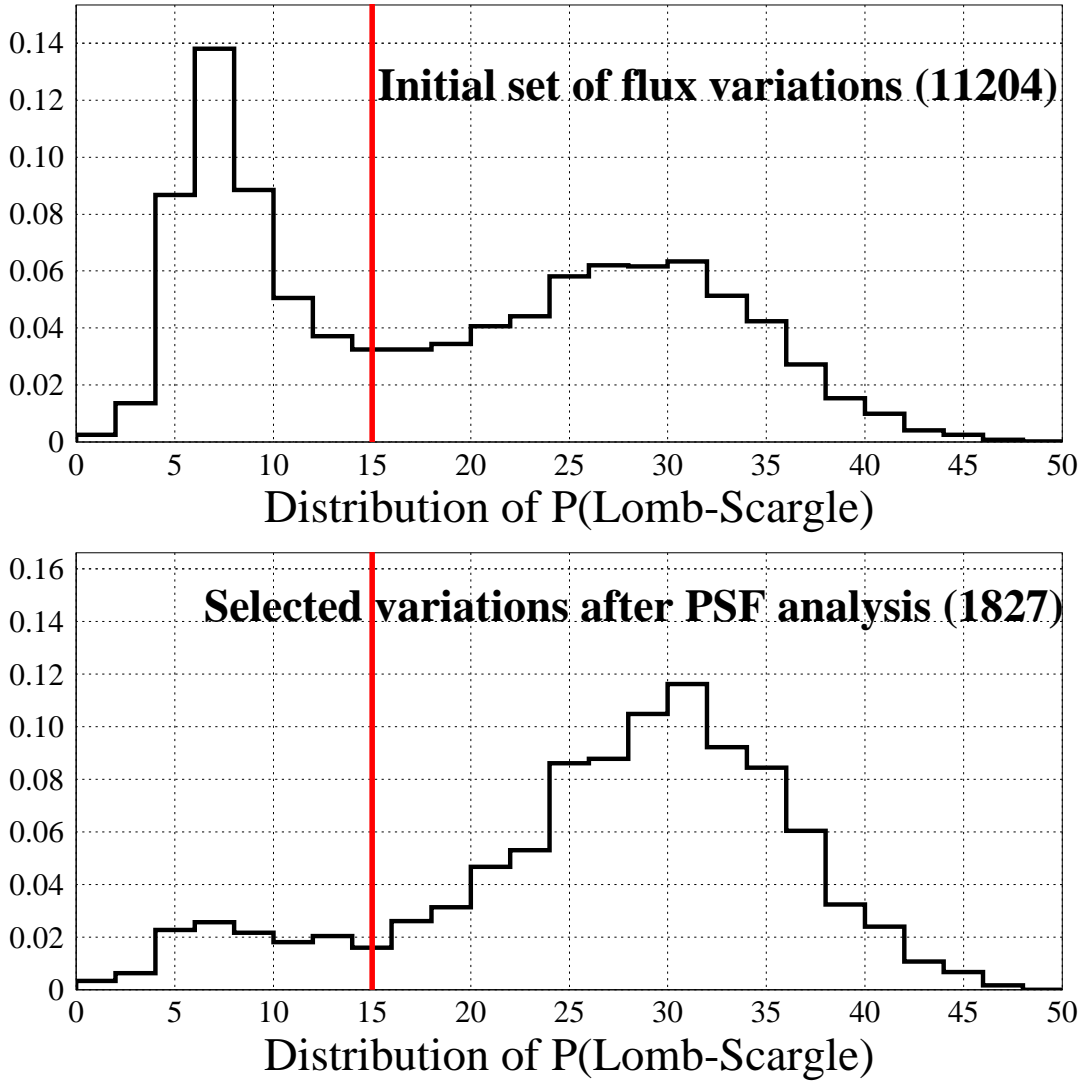}
\caption{
Normalized distribution for the ``power'', $P$,
of the Lomb-Scargle periodogram analysis on the 
extension of our selected flux variations
on the corresponding INT light curves.
Top panel: the full set of initial selected light curves.
Bottom panel: the surviving set of 
selected light curves after the PSF analysis
(see text for details). The vertical solid
line at $P=15$ indicates the threshold value
used within our selection pipeline.
\label{fig:lomb}
}
\end{figure}

The pipeline we use closely follows that described in \cite{novati09} which
we refer to for full details.  Hereafter we highlight the relevant steps.
We work in the pixel lensing regime so that the search for flux variations is carried out
along all the pixels of the image. The analysis is carried out
on OAB data working on each year separately. First, we have to establish a list of
flux variations. To this purpose we select light curves showing
at least 3 consecutive points (one per night) above the baseline level
at 3 sigma level, in both $R$ and $I$ band and then ask for the threshold cut,
for $R$ band data, $Q>50$, where $Q$ is the ratio of the $\chi^2$ of a straight line
over a Paczy\'nski  fit \citep{novati02,novati03}. We recall that each flux variation is spread over
a full cluster of nearby pixels whose identification therefore requires an 
analysis based on the  spatial information of the images \citep{novati02,novati09}.
This way we select a first sample of some 11204 flux variations. 
As a second step we want to remove all spurious variations (bad pixels, cosmic rays,
variations induced by the seeing and so on). 
To this purpose we study the shape of PSF of the bump on a difference image obtained selecting
images at the peak and images along the baseline having similar seeing conditions \citep{novati09}.
This way the number of selected flux variations drops to 1827. Next, we test the stability 
of the baseline, as indeed we expect most of these variations to be intrinsic variable signals.
To this purpose we carry out a Lomb-Scargle 
periodogram analysis \citep{lomb76,scargle82} that we implement
following \cite{numrec92} along three years of INT data  
\citep{point01,paulin03} and consider as a statistics the associated power $P$.
As a threshold value to distinguish between noise and signal we use $P=15$ (Fig.~\ref{fig:lomb}). 
(As a test that our initial set of 11204 variations is indeed dominated by spurious signals,
therefore with most chances to show a stable baseline along the corresponding INT light curve,
we find that most of the variations excluded  with the PSF analysis have in fact $P<15$.)
Experience (and superpixel photometry) teach us however that this way we may lose bona fide
microlensing candidates whose light curve may be superimposed on a (possibly nearby)
variable (as the POINT-AGAPE PA-N1 event, \citealt{point01}). We therefore 
allow for flux variations with a variable baseline provided that the flux difference 
on OAB data be significantly larger than the corresponding one on INT data.
The set of flux variations then reduces to 612. 
Finally we adopt three further selection criteria
to constrain the shape of the light curve:
the first one for good enough sampling along the flux variations \citep{novati09};
the second for compatibility with Paczy\'nski, testing the reduced $\chi^2$
and asking $\chi^2 < 10$; the last one for large enough variations, 
with a threshold on the flux difference at maximum magnification expressed
in term of magnitude $\Delta R_\mathrm{max}<21.5$.
This way we are left with 4 microlensing candidate events:
OAB-07-N1 and OAB-07-N2 already selected and presented in
\cite{novati09}, with the second further discussed in \cite{novati10};
a candidate out of 2008 data, further discussed below and finally eliminated
from the selection; and, out of the 2010 data, a microlensing candidate already reported
by PAndromeda, PAnd-4 \citep{lee12}, that we may also dub OAB-10-S3
(N and S stand for North and South, the OAB field where the candidate is located).

\begin{figure}
\epsscale{1}
\plotone{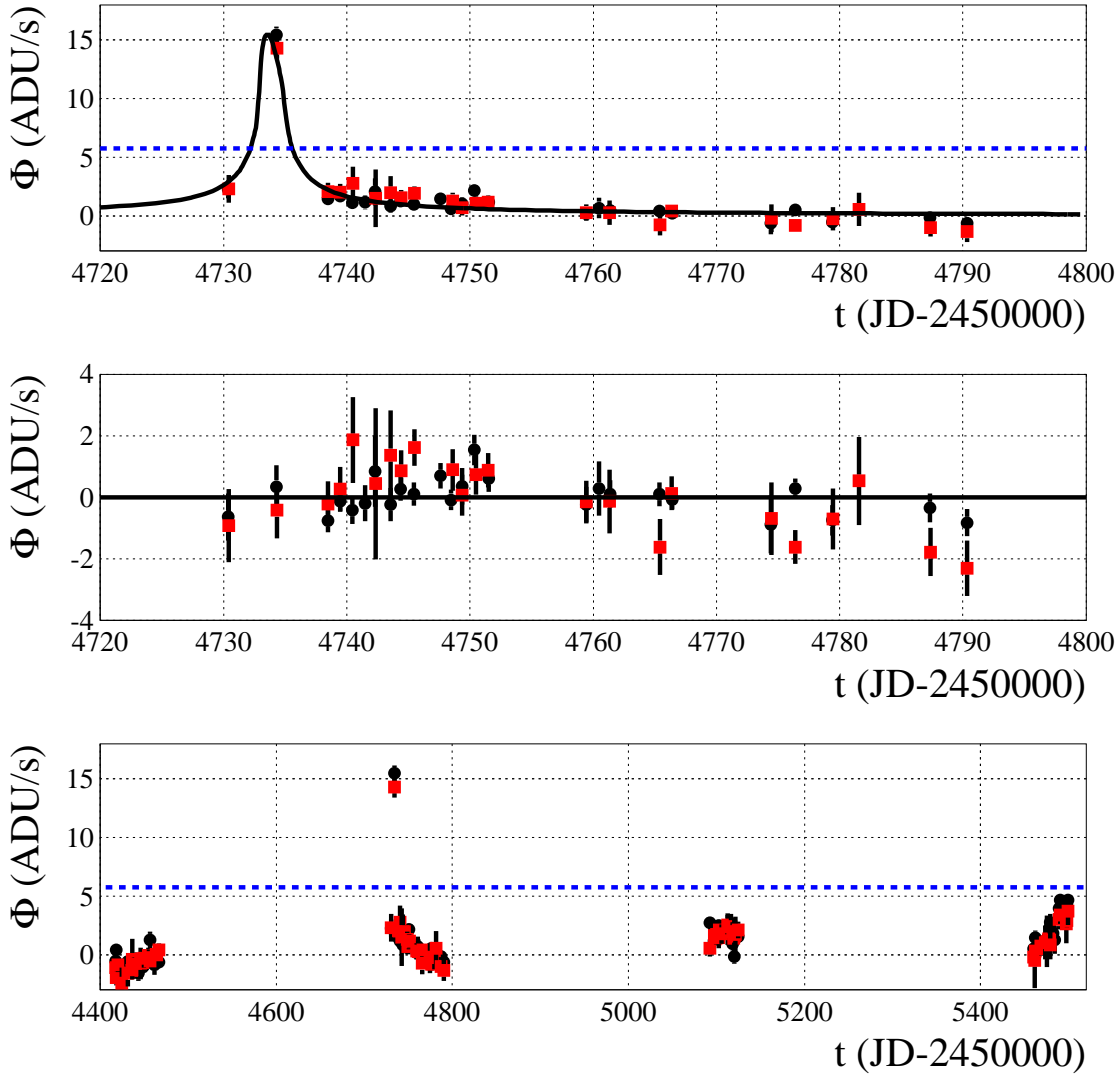}
\caption{
The flux variation detected in 2008 on OAB data
selected by our automated selection pipeline
but finally eliminated from the sample
of microlensing candidates (see text for details).
Top panel: OAB 2008 light curve with
the solid line represents the best Paczy\'nski fit
to the data. The dashed horizontal line indicate the flux 
difference with respect to the baseline corresponding
to the flux deviation at maximum of the underlying variable 
as analysed along the 1999-2001 INT data (see text for details).
Middle panel: OAB 2008 light curve of the residuals with respect
to the best Paczy\'nski fit. Bottom panel: OAB 2007-2010 light curve and dashed
horizontal line as in top panel.
All panels: $R$ and $I$ band (rescaled) data 
are shown with circle and square symbols, respectively. 
The flux values on the $y$-axes are rescaled with respect to the $R$-band values.
\label{fig:oab08xx}
}
\end{figure}

The 2008 selected flux variation, in 
$\alpha,\delta=  0^\mathrm{h} 42^\mathrm{m} 49.22^\mathrm{s}, 
41^\circ 22' 42.5''$ (J2000.0) 
at a distance of $6.6'$ from the M31 center, 
with maximum magnification at  4734. (JD-2450000.0),
has a very short, half-width-half-maximum
duration, $t_\mathrm{FWHM}$, below 3 days, 
and a quite bright bump, with flux difference
at maximum magnification expressed in term of magnitude
$\Delta R_\mathrm{max}\sim 20.0$ and color $R-I\sim 0.9$
(at the observed peak), Fig.~\ref{fig:oab08xx}. On the other hand,
the corresponding extension along the INT light curve show a clear variable signal
(with $P=31$). Indeed, also OAB data (although penalised by a shorter baseline per year of data),
in 2008 as well along the full four years of data, show evidence of that variation.
A closer astrometry inspection, with rms of the relative OAB-INT
astrometry below 0.2 OAB pixel level, reveals that the INT variable
sits some 4 INT pixels away from the pixel corresponding
to the OAB variations, in a position that coincides
with that of the variable identified also on OAB data, 2 OAB pixel away
from that of the candidate (OAB and INT pixels
cover $0.33''$ and $0.58''$ respectively, for a distance
of the candidate from the underlying variable of about $1.2''$).
The selected flux variation is definitely on a different
position with respect to the underlying variable
running along the same superpixel light curve of the microlensing candidate.
From INT data we infer the color of the variable as $R-I=1.1$,
somewhat redder than the OAB variation, 
with peak magnitude $R=21.2$, more than 1 magnitude fainter
than the OAB variation.
As apparent from inspection of the OAB light
curve, the sampling along the bump is poor, with a single data point 
(in both $R$ and $I$ data, with five
images of that night per filter all clearly showing the variation, 
and with no indications of any trend during the night)
well above of the variable baseline
and no data, because of bad weather, on the 3 nights immediately before and after the peak.
Additionally, a comparison of the two OAB light curves, that centered on the candidate
and that centered on the position of the variable, strongly suggest that 
the flux excess with respect to the baseline for the data points
immediately before and after the peak,
at the origin of the initial trigger of this flux variation
within the selection pipeline,  should be attributed
to the underlying variable rather than to the candidate
which therefore is left with a single significant data point
(per band) along the bump. As an initial threshold
we ask for three consecutive points, in each band,
above the baseline level at 3 sigma level, 
we are bound to exclude this flux variation,
which our available sampling do not enable us
to properly characterize, from our selection.

\begin{figure}
\epsscale{1}
\plotone{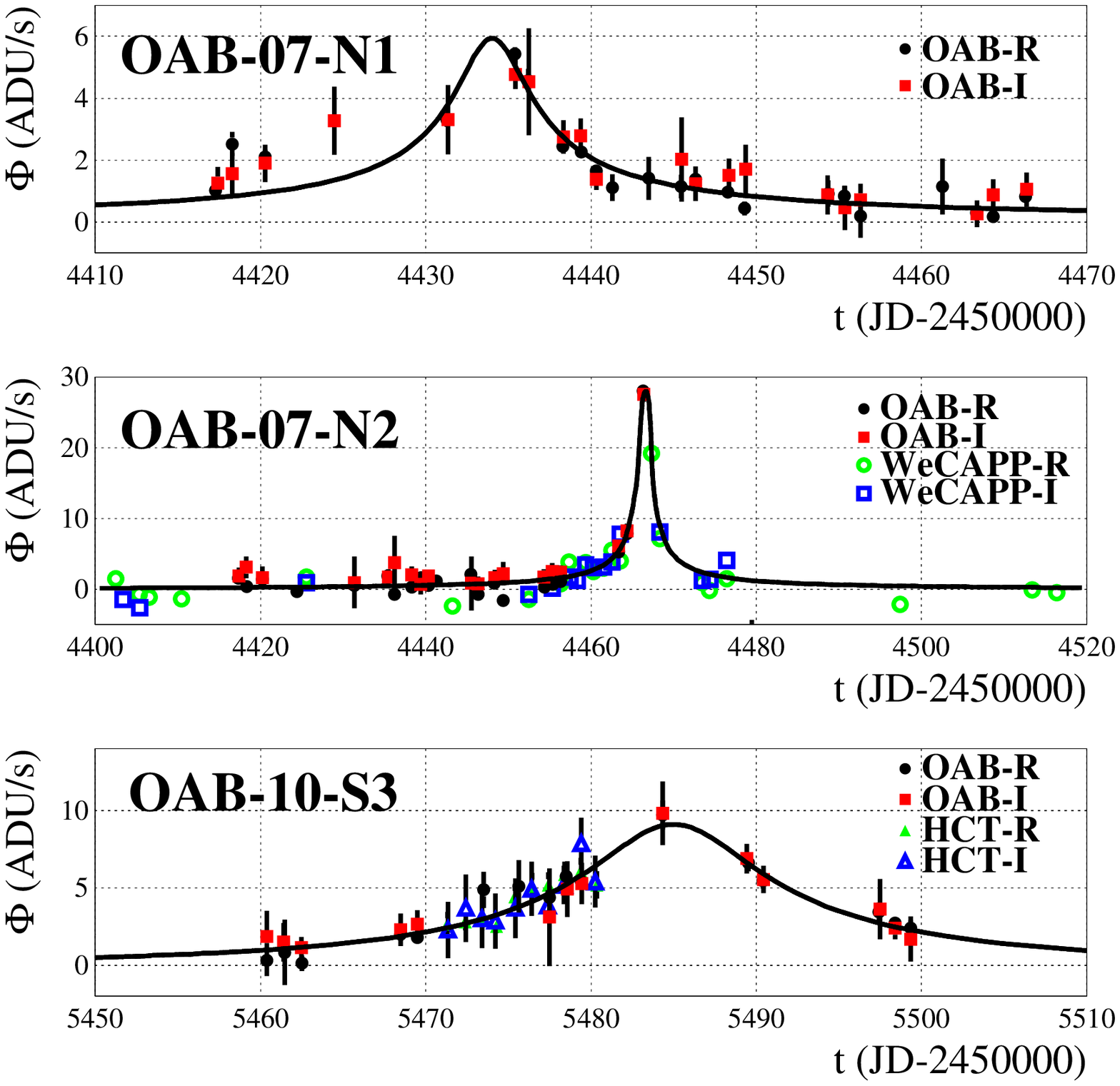}
\caption{
Light curves for the three selected microlensing candidate events
of our automated selection pipeline.
From top to bottom panel: OAB-07-N1 and OAB-07-N2,
both first presented in \cite{novati09},
and OAB-10-S3 first published as PAnd-4 in 
\cite{lee12}. The solid line represents the best Paczy\'nski fit
to the data. Middle and bottom panels:
besides the OAB data (filled symbols) we report
the additional WeCAPP data also used for
the analysis carried out in \cite{novati10};
the additional HCT data of our 2010 campaign.
The flux values on the $y$-axes are rescaled with respect to the OAB $R$-band values
expressed in ADU/s.
\label{fig:gold3}
}
\end{figure}

\begin{center}
\begin{deluxetable}{rrrr}
\tablecolumns{4}
\tablewidth{0pc}
\tablecaption{Main characteristics of the three candidate microlensing
events selected within the fully automated 2007-2010 OAB selection pipeline.
\label{tab:nobs}}
\tablehead{
\colhead{} & \colhead{OAB-07-N1} & \colhead{OAB-07-N2} & \colhead{OAB-10-S3}\\
}
\startdata
$\alpha$ (J2000) &
$0^\mathrm{h} 42^\mathrm{m} 56.70^\mathrm{s}$ &
$0^\mathrm{h} 42^\mathrm{m} 50.36^\mathrm{s}$ & 
$0^\mathrm{h} 43^\mathrm{m} 11.52^\mathrm{s}$\\
$\delta$ (J2000) & 
$41^\circ 22' 49.8''$ &
$41^\circ 18' 40.1''$ &
$41^\circ 13' 20.2''$ \\
$\Delta_{\mathrm{M}31}$ (arcmin) & 7.1 & 2.8 & 5.8\\
\hline
$t_0$ (JD-2450000.0) & $4434.0\pm 0.2$ & $4466.1\pm 0.1$ & $5485.1\pm 0.3$\\
$t_\mathrm{FWHM}$ (days) & $7.7\pm 0.7$ & $1.4\pm 0.3$ & $15.\pm 2.$ \\
$\Delta R_\mathrm{max}$ & $21.1\pm 0.1$ & $19.2\pm 0.2$ & $20.6\pm 0.1$\\
$R-I$ & $1.2\pm 0.1$& $1.2\pm 0.3$ & $0.8\pm 0.1$\\
\enddata
\tablecomments{
$t_\mathrm{FWHM}$ and $\Delta R_\mathrm{max}$ are the duration and the flux
difference from the baseline level expressed in term
of magnitude according to the fit scheme of \cite{gould96}.
For OAB-07-N1 and OAB-07-N2 the results are slightly different,
still compatible within errors, from those reported in \cite{novati09,novati10}
because of the extended baseline. For OAB-07-N2
we report the results of the fit along the joint data sets OAB plus WeCAPP. 
For OAB-10-S3 we include within the fit the data acquired during our 2010 HCT campaign.
OAB-10-S3 has been first published as PAnd-4 in \cite{lee12}
and we are unable to explain the rather large difference
for some of the reported values with respect to those reported in their
Tables~3 and 4.
}
\end{deluxetable}
\end{center}

The remaining 3 candidates, on the other hand, all show
a stable INT extension light curve (Lomb-Scargle power $P<7$ for all three of them)
as well as a flat baseline on the OAB data on the years off bump.
They are  further discussed in their respective discovery
papers, here we report their main characteristics, Table~\ref{tab:nobs},
and show their light curves, Fig.~\ref{fig:gold3}.

For the 2010 season we have at our disposal also the additional HCT data set,
with a smaller field of view than the OAB one (recovering a fraction of the area of about 60\%) 
and sampled  along a 10 consecutive days baseline, about 1/3 of the 
overall baseline of the OAB 2010 season. PAnd-4/OAB-10-S3 lies within the HCT field of view.
However, the last HCT data point falls 4~nights before the peak of the event, still, 
as the event is rather long the HCT data
are still useful as they cover (and nicely overalp with the OAB data, 
Fig.~\ref{fig:gold3}, bottom panel)
the rising part and help us to better constrain the event lensing parameters.

As an additional analysis we search for X-ray counterparts
of our candidates on archive data. A positive match with
a known X-ray source may indeed be
an hint of a possible non-microlensing origin of the
corresponding optical flux variation. 
In particular, we cross match our data with both the
``M31 Deep \emph{XMM}-Newton Survey X-Ray Source Catalog'' \citep{stiele11},
an updated version with respect to that we used in our previous
analysis \citep{novati09}, and the Chandra analysis
``LMXBs in the bulge of M31'' \citep{voss07}.
As for the astrometric precision
\cite{stiele11} report a $3\sigma$ positional error for every entry,
statistical and systematic, usually around a few arcsec;
for the Chandra analysis \cite{voss07} report
an indicative range of values, from $0.1''$ to $0.4''$,
depending on the brightness of the sources; finally, our astrometric
solution is done using about 360 bright stars per field
cross-identified with sources in \cite{massey06} with (statistical)  rms below $0.3''$. 
The nearest X-ray source to one of our candidates
is that lying at $25''$ from OAB-07-N2
(in the Chandra catalog, the nearest in the 
\emph{XMM}-Newton one is reported at a distance of $27.2''$
with positional error of $1.84''$).
For the given errors we can safely rule out an identification.  
The same conclusion applies, a fortiori,
for both OAB-07-N1 and OAB-10-S3.
For the first, the nearest source is found at more than $1'$ (both catalogs);
for the second at $35.1''$ (Chandra) and $44.4''$ with positional
error of $1.83''$ (\emph{XMM}-Newton).
The situation is less clear for the 2008 flux variation
we have eliminated from our analysis. 
A positive cross-match of the positions looks plausible
if we consider the \emph{XMM}-Newton data,
with a source at $5.9''$ with   positional error of $4.75''$.
This does not hold  any more, however, both because of the smaller 
positional error and the increased distance to our flux variation
($7.5''$), when we look at the Chandra data.

Finally, we may wonder, faced with the large possible variety of microlensing
signals (binary lenses and so on), about the impact on our analysis of the
requirement of  compatibility with a Paczy\'nski shape.
If we drop this requirement, we end up with only 2 additional selected flux
variations. The first is a clear variable signal, for which we
have no available INT data but whose nature is revealed by the analysis of
the light curve in the OAB data along the years off bump.
The second is a more interesting case: a rather blue, $R-I\sim 0.3$, 
extremely bright, $\Delta R_\mathrm{max} \sim 17.6$, variation,
located at $\alpha,\delta = 0^\mathrm{h} 43^\mathrm{m} 02.44^\mathrm{s}, 
41^\circ 14' 10.3''$ (J2000.0), at $3.9'$ from the M31 center, and occurring in 2008 around  
4760. (JD-2450000.0) showing several peaks along a time scale
of a few days as well as 
signs of chromaticity. We tried a binary lens fit \citep{bozza10} 
on this flux variation but we could not find any viable solution.
Therefore, we attribute this flux variation to some kind of unidentified
cataclysmic variable. As above, we have checked for possible
X-ray counterparts. The nearest (bright) source, both the \emph{XMM}-Newton
and the Chandra catalogs, lies at about $25''$ and
should therefore, for the given errors, be unrelated to this flux variation.

\cite{lee12} presented the results of the 2010 season
of their PAndromeda (M31) pixel lensing survey.
They used a 1.8m telescope with a very wide field of view instrument (7~deg$^2$)
and obtained 91~nights of data along about 5~months
in two broad band $r_\mathrm{P1}$ and $i_\mathrm{P1}$ filters.
In particular they presented results for a search
of microlensing events within the inner $40'\times 40'$
of M31. Overall, they reported six microlensing event candidates.
We can use the 2010 PAndromeda results to test our OAB 2010 pipeline.
In spite of the very large ratio of ours and
the PAndromeda monitored field of view, 
about 20\% considering only the (small)
fraction of the overall field of view
on which \cite{lee12} carried out their
microlensing search, as a consequence
in fact of the sharp decrease of the expected signal
moving outwards from the M31 center,
four out of the six PAndromeda candidates
falls within the OAB fields of  view (PAnd-1,2,3,4).
Because of the much longer 2010 PAndromeda baseline, 
however, only two of these have been detected 
in October while the OAB campaign was going on (PAnd-1 and 4).
As discussed, we find PAnd-4 to coincide with OAB-10-S3 also detected
within our pipeline. PAnd-1, which has a very short
duration, $t_\mathrm{FWHM}=3.1~\mathrm{d}$ \citep{lee12},
unfortunately falls within a gap of the OAB sampling.
On the OAB data, along its (short) bump we detect two points,
well above $3\sigma$ level of the baseline, 
according to our selection, however, clearly insufficient to characterize, 
if not even to trigger, a detection.
The HCT data span exactly the moment of the PAnd-1 peak,
unfortunately, however, PAnd-1 is not included within the HCT
field of view. We may therefore conclude that the
output of ours and the PAndromeda pipeline are compatible.
We consider this conclusion to strengthen the results of the OAB
pipeline also for the previous years.

Single bump, achromatic, suitably sampled with large enough S/N, Paczy\'nski-like flux variations
can be considered reliable microlensing \emph{candidates}. Excluding binary lenses and/or similar
cases where the intrinsic microlensing nature of the event can be accepted beyond any doubt,
these flux variations are bound to remain within this limbo. A still possible
background is that of cataclysmic variables, which are usually single bump flux variations (at least
within the time scale of the usual considered duration for the analysis of the baseline stability).
However, these are usually bluer than the typical  M31 microlensing candidates, and in particular
of those discussed here, and tend to show, as is typical for intrinsic variables, an asymmetry
along the flux variation with a sharper rising part. For the case under examination, the intrinsic 
microlensing nature of two of the reported flux variations is further supported by additional data
by WeCAPP, for OAB-07-N2, and HCT (presented here) and PAndromeda for OAB-10-S3.
Indeed, the simultaneous detection on multiple pipelines and/or multiple 
data sets of the same flux variation, 
even if by itself can not be taken as a proof of the genuine microlensing 
nature of the flux variation, may  
make us more confident on its interpretation. 
This is for two main reasons. First, the joint analysis with additional data may further constrain 
the microlensing parameter space. Second, each pipeline (a fortiori  
with a different data set), 
in its broadest sense (data reduction, photometric analysis, 
flux variation search and characterization), comes 
with its own systematics which tend to 
be ruled out by multiple detections. 
More specifically, in \cite{novati09} with OAB data alone  
OAB-07-N2 was not fully sampled and we could only
put forward a guess on its microlensing nature.
The joint analysis with the additional WeCAPP data 
then enabled us \citep{novati10} to probe the symmetric and achromatic
shape of the full flux variation then confirming
the microlensing interpretation. Furthermore,
the dense sampling made possible a much more refined analysis
of the microlensing parameter space.
Indeed, together with an additional analysis
on the underlying source flux on archival data,
the joint OAB plus WeCAPP light curve 
enabled us to conclude, even if only marginally, through a
study of the lens proper motion \citep{gould94,hangould96},
in favor of the MACHO nature of the lens.
As for OAB-10-S3, the HCT data presented here, even if not
necessary to enhance its detection,
enabled us a better characterization of the microlensing parameters.
This flux variation enjoys then of being 
selected as a microlensing candidate
by two completely independent pipelines
(on different data sets), specifically,
besides the present one, also
by PAndromeda as PAnd-4 \citep{lee12}.
For purposes of the following analysis we therefore will consider the
three flux variations selected within our pipeline as \emph{bona fide} microlensing variations.

\subsection{Monte Carlo and efficiency analyses} \label{sec:mls}

For the analysis of the expected signal we closely follow the scheme outlined
in \cite{novati09} and references therein  which we refer to for full details.
First, we build a Monte Carlo simulation (based on the original work
in \cite{agape93,agape97}) where, 
on top of the astrophysical model of all the quantities of interest
we simulate the microlensing flux variations.
The evaluation of the expected signal for a microlensing
experiment is based of the microlensing \emph{rate} \citep{griest91,novati08},
with the specific case of M31 pixel lensing also discussed in 
\cite{han96,baltz00,gyuk_crotts00,kerins01,riffeser06}.
Our model of M31 is based on the \cite{kent89} data.
For both M31 bulge and disk stars, we make use of a synthetic luminosity 
function extracted from IAC-star \citep{iac04} with sources
expected up to a magnitude of roughly $M_I\sim 2$.
Finally, the flux variations are simulated 
as single-lens microlensing events accounting for finite source size 
\citep{witt_mao94} and we reproduce the observational conditions,
in particular the sampling, of our OAB campaign. 
Within the Monte Carlo we carry out a first (knowingly over-optimistic)
selection pipeline asking for the flux variations
to have at least 3 consecutive points 3 sigma above
the baseline level. Monte Carlo selected light curves
may however not be selected within our data set. 
Within the Monte Carlo we can not in particular
reproduce those steps of our pipeline where the  spatial information across 
the images comes into play: the cluster analysis
we carry out to identify the initial set of flux variations and
the PSF analysis we use to exclude spurious signals;
additionally, within the Monte Carlo we do not
reproduce all the problems intrinsic to the images
such as crowding, background flux variations due
to underlying variables and so on: all these
aspects must however be taken into account.
To get to a  reliable estimate of the expected signal
we therefore inject, making use of the \texttt{daophot} tasks
within IRAF, (part) of these Monte Carlo selected light curves
on the real data ($R$ band only), just after the basic CCD reduction,
and then run our selection pipeline from scratch.
Finally, as a result, we end up with  the distributions of the parameters and the number
of events for the expected signal. As lens populations we consider M31 bulge and lens
stars (``self-lensing'') and MACHO lensing as MACHOs belonging to the M31 and MW halos, 
for which we study a set of delta mass functions within the range
$10^{-3}-1~\mathrm{M}_\odot$. Specifically, to minimize the statistical noise
on the number of expected events, within the Monte Carlo we simulate $10^9$ events for self lensing
and as much for each mass of MACHO lensing, per year. Out of the selected events
within the Monte Carlo we then inject 12000 events for self lensing
and for each MACHO mass value per year (to avoid overlaps among the injected
events on the images we split the analysis so to have 500 events per field for each run).

\subsection{The statistics on MACHOs} \label{sec:machos}

\begin{figure}
\epsscale{1}
\plotone{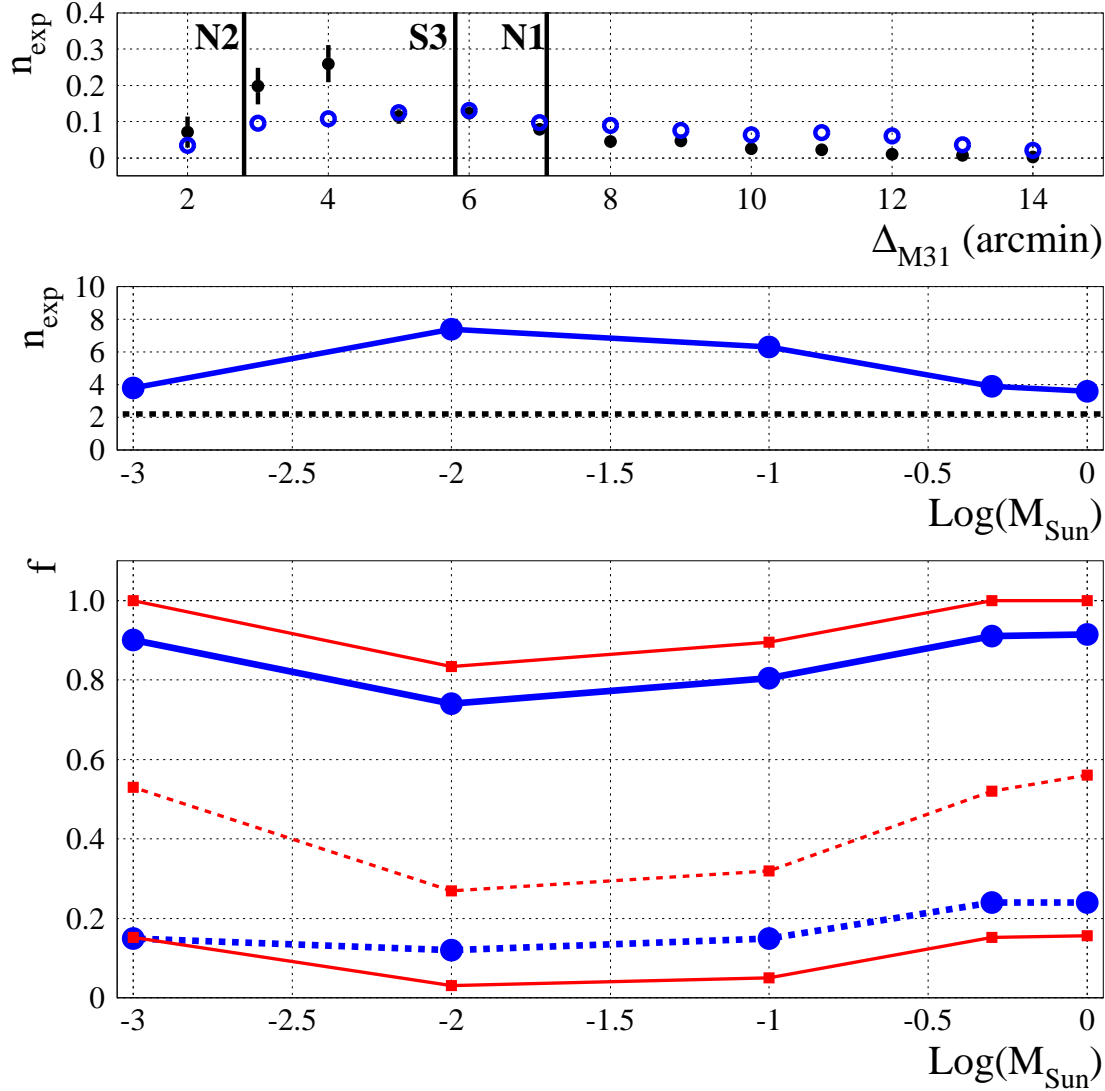}
\caption{
Top panel: the (normalized) expected number of events
for self lensing (filled symbols) and MACHO lensing
as a function of the distance to the M31 center (arcmin).
The (statistical) error bars appear only when they
exceed the size of the symbols. The solid vertical lines
indicates the position within this parameter space
of the selected candidates.
Middle panel: the number of expected MACHO lensing events,
for full (M31 and MW) halos ($f=1$) as a function
of the (logarithm of the) MACHO mass; the horizontal dashed line
indicates the expected number of self-lensing events. 
Bottom panel: limits on the halo mass fraction
in form of MACHOs, $f$, as a function of the (logarithm of the)  MACHO mass
obtained from a Bayesian analysis based on the evaluation
of the likelihood comparing the expected, MACHO and self lensing, signal
to the 3 observed events (see text for details).
Circle symbols: modal values (dashed lines) and upper 95\% CL limit in the case where
no a priori hypotheses are made on the nature of the observed events.
Square symbols (and thin lines): modal values (dashed lines) and upper and lower 95\% CL limits
obtained under the additional hypothesis that, following
the analysis in \cite{novati10}, OAB-07-N2
\emph{is} a MACHO lensing event.
\label{fig:res}
}
\end{figure}

The driving astrophysical question of the present analysis
is the content in MACHOs of galactic halos. It is therefore
of primarily interest to address the issue of the \emph{nature}
of the observed events, whether self lensing or MACHO.
Besides the already discussed and peculiar case of OAB-07-N2,
the main statistics at our disposal, also considering
the power of investigation within the detection efficiency analysis,
are the magnitude at maximum and the color, which at most can be used
to assess the coherence of the analysis
with the expected signal with no reference however
to the specific lens population, the duration and the position.
As for the duration, $t_\mathrm{FWHM}$, we recall from previous
analyses \citep{kerins01,novati05,riffeser06}, and specifically
for the OAB data our analysis in \cite{novati09}, that most events are expected,
as indeed we find, to last less than about 10-20 days,
but this parameter is unfit to distinguish among the lens populations,
at least for those of roughly equal lens mass (thus, specifically, for stellar
lenses and MACHOs around $0.5~\mathrm{M}_\odot$)
for which one would rather need a reliable estimate of the Einstein time.
On the other hand, the small number of events at our disposal makes the position of the events 
of limited use to this purpose, moreover it should be recalled
that the analysis of the originally proposed signature
of M31 MACHO lensing versus self lensing due to the M31 disk inclination may in fact
be complicated by effects of differential extinction \citep{an04,montalto09}.
Still, self-lensing events tend to be more clustered than MACHOs 
around the M31 center, and specifically this holds
for our experimental configuration \citep{novati09}.
In fact, this is the underlying motivation, in the following likelihood
analysis, for binning the space in unit of distance from the M31 center
(Fig.~\ref{fig:res}, top panel).
The main statistics to disentangle MACHO and self lensing signals 
which is left is therefore the expected \emph{number} of events. The results
of our analysis, Monte Carlo simulation followed by
the efficiency analysis, are as follows.
For self lensing we expect $2.2\pm 0.2$ events.
For MACHO lensing, for full ($f$=1) M31 and MW halos,
from about 4 up to above 7 events, according to the mass value
(Fig.~\ref{fig:res}, middle panel)
and with statistical relative error  around 10\%,
with a maximum for the expected signal at $10^{-2}~\mathrm{M}_\odot$ before 
a decrease, for $10^{-3}~\mathrm{M}_\odot$,  due to the drop
of sensitivity for very short duration events related
to our (insufficient) sampling.
The reported \emph{statistical} error is  that associated to the Monte Carlo simulation,
for which the (Poisson) error scales as the square root of the simulated events
(Section 5.1 of \citealt{novati05}) with the final budget dominated by large 
by the error on the efficiency.
Next, we may compare this result for the expected signal with $n_\mathrm{obs}=3$ observed events
with the additional information, as already recalled,
that OAB-07-N2 may indeed be rather a MACHO than self lensing.
Leaving aside this information for the moment, the conclusion
of the analysis, based on the bare numbers, is that self lensing is fully able to explain
the observed rate. Specifically, according to the Poisson statistics
followed by the number of events, for the given expected mean signal
(and excluding for a moment the MACHO signal),
the 95\% CL upper limit is for 6.3 events, well above the observed rate.
We have then to consider the expected MACHO lensing.
Here a fundamental remark is that the expected
signal, for \emph{full} halos, is not much larger than the self lensing one. Together, these results
suggest that the current experiment is unable to set lower limits to MACHOs
and to establish, if any, only  rather weak upper limits.
On the other hand, that same smallness of the expected MACHO lensing signal renders,
as soon as we acknowledge the MACHO nature of even only one event, as we may do for OAB-07-N2 
according to the analysis in \cite{novati10}, the situation altogether different as we may expect
at once to find a sizeable \emph{lower} limit for the halo mass fraction in form of MACHOs.

We can quantify the above statements by inferring from the data a probability distribution for
$f$ through a Bayesian analysis based first on the evaluation of the likelihood function.
To this purpose we closely follow the approach outlined in \cite{novati05} (for a specific discussion
on the use of the likelihood within a microlensing analysis we also refer to \citealt{novati13}). 
We bin the entire field of view  into $N_\mathrm{bin}$ bins, for each we have the model
prediction $x_i$ ($i=1,\ldots,N_\mathrm{bin}$) and the observed number of events $n_i$.
For fixed model, the different $x_i$ are not independent as they all depend on the halo fraction, $f$,
\begin{equation} \label{eq:signal}
x_i\left(f,m\right)=b_i+f s_i\left(f=1,m\right)\,,
\end{equation}
with $m$ being the MACHO mass and
where the \emph{signal} we look for, $s$, is the number
of MACHO events, the \emph{background}, $b$, being the self lensing signal
(as our purpose is to constrain $f$ we may
rather refer to the signal as the product $f\cdot s$).
The likelihood is the product of the individual 
probabilities of the $n_i$ in each bin
\begin{equation} \label{eq:like}
L\left(n_i|x_i\right) = \prod_{i=1}^{N_\mathrm{bin}} \frac{1}{n_i!} \mathrm{e}^{-x_i} x_i^{n_i}=
\mathrm{e}^{-N_\mathrm{exp}} \prod_{i=1}^{N_\mathrm{obs}} x_i\,,
\end{equation}
where in the second step we have specialised the bins
to contain either 0 or 1 observed event \citep{gould03}.
As previously addressed, for binning, we choose the 
distance from the M31 center, and we use $1'$ bins.
Keeping the MACHO mass fixed as a parameter,
for a flat prior for $f$ different from zero in the interval $(0,1)$,
given the likelihood we can then evaluate the probability distribution $P(f)$.
We evaluate the modal value for $f$ and,
around it, the 95\% CL region which then
defines the lower and the upper limits. The results
of this analysis are shown in Fig.~\ref{fig:res}, bottom panel.
If we do not introduce any prior within the analysis, as anticipated, the observed
signal turn out to be compatible with the 
expected self lensing rate, with no lower limits 
for $f$ and upper limits above 70\% (90\% in the range
0.5-1~M$_\odot$). If instead we impose that OAB-07-N2
\emph{is} a MACHO, namely within Eq.~\ref{eq:signal}
we set the self-lensing background to zero, $b=0$,
we find a, sizeable, 15\% \emph{lower} limit
for $f$ in the mass range 0.5-1~M$_\odot$,
with no upper limit and somewhat smaller limits
moving down to $10^{-2}\mathrm{M}_\odot$.
Finally, the observed rate (3 events)
is in fact also compatible with no self lensing
(for an expected signal of 2.2 events).
If we assume this extreme case as a working hypothesis
the lower limit for $f$ in the mass range 0.5-1~M$_\odot$
would rise to about 30\%.

\section{Conclusion} \label{sec:end}

We have presented the final
analysis of the 4-years, 2007-2010, pixel lensing
campaign of the PLAN collaboration towards M31
aimed at the search and the characterization
of microlensing events. The driving scientific motivation
is the search for a dark matter signal in galactic halos
in form of compact objects, MACHOs. The specific aim
of the campaign is to better understand
the signal coming from the putative MACHO
population as compared to the background signal
of self-lensing events, defined as opposed
to MACHO lensing with the lens belonging
to known stellar (M31) populations.
To this purpose we monitored the central region
of M31, where the expected rate is larger for both signals,
still with self lensing expected to be more clustered
around the M31 center, looking for new
microlensing events. A key aspect of our analysis
is the use  of a full automated pipeline
for the search of microlensing-like flux variations
which, besides leading us to the determination 
of a set of microlensing candidates,
enables us to reliably estimate the detection efficiency. 
For a given astrophysical model this eventually
enables us to reliably estimate  the expected signal
through a Monte Carlo simulation of the experiment.

The analysis is based on data collected at the 1.5m telescope
of the Astronomical Observatory of Bologna (OAB)
in two broad $R$ and $I$-bands
in two fields $13'\times 12.7'$ around the M31 center.
After a first year pilot season (2006) \citep{novati07}
the observational campaign eventually lasted
four years (2007-2010) with an awarded
baseline  to our survey, on average,
of some 48 night/year plus, in 2010,
10 consecutive nights of complementary data from
the 2m Himalayan Chandra Telescope (HCT). 
Altogether, however, bad weather and/or generally
unsuitable observational conditions 
introduced several gaps within our sampling
with the final analysis based on the data
collected, overall, during about 90 nights 
(excluding the contribution of HCT data).
The expected short duration of the microlensing
flux variations together with the small rate
of events magnify the impact of this problem.
Indeed, this is made explicit also from the analysis presented
in this paper. We have discussed
the case of a flux variation preliminarly
selected, in the 2008 season, but finally rejected
because of incomplete sampling along the bump.
For the same reason we did not select
the microlensing candidate PAnd-1 \citep{lee12}
even if included within our field of view and baseline.

The results of the pipeline are as follows.
Overall, we select three microlensing candidate events:
OAB-07-N1 and OAB-07-N2, both first presented 
in \cite{novati09}, and a third candidate occurred
during the 2010 season and already published by
PAndromeda in \cite{lee12} as PAnd-4, which we also dub OAB-10-S3,
and for which we also have data from HCT.
The results of our pipeline for the 2010 season turns out to be compatible
with those of PAndromeda \citep{lee12}
(thus strengthening its conclusions also  for the previous seasons).
As discussed within the text, the detection of the same
flux variation on multiple pipelines and/or data sets
is useful for the purpose of its interpretation
as a microlensing candidate. First, additional data may help
to better constrain the candidate microlensing
parameter space. Second, any independent detection comes
with the bonus of removing, 
if any, the systematics of each pipeline.
Besides the case of OAB-10-S3 (PAnd-4) we recall OAB-07-N2, first
presented in \cite{novati09}, of which we could then perform 
a new analysis thanks to additional WeCAPP data \citep{novati10}.
In particular, these made possible a refined analysis of the lensing parameter
space, specifically of the lens proper motion, 
which enabled us to conclude in favor
of the MACHO nature of the lens.
For purposes of the analysis we consider all three candidates
as \emph{bona fide} microlensing variations.

The observed rate, based on the \emph{number} of events,
is compatible with the  expected self-lensing signal.
A major outcome of our analysis, though,
is that the expected MACHO lensing, for full M31 and MW halos, is only
marginally larger than self lensing, which is a different situation from analyses
towards the Magellanic Clouds where the expected
self-lensing signal is evaluated to be much smaller
than that of MACHO lensing. This result, together
with our small statistics of events at disposal,
prevents us from drawing strong constraints on the  putative MACHO population.
This situation makes extremely important, whenever
possible, the detailed analysis of single events addressing the issue of their nature,
as was the case for the event  POINT-AGAPE-S3/WeCAPP-GL1 \citep{riffeser08},
and as we could do for OAB-07-N2 \citep{novati10}.
Indeed, the hypothesis on the MACHO nature of OAB-07-N2, as suggested by that last analysis, drives 
a sizeable lower limit for the halo mass fraction in form of MACHOs, $f$. Quantitatively, we evaluate
an expected self lensing signal of 2.2 events,
fully compatible therefore with our 3 observed events,
and a MACHO lensing, for full halos, of 4-7 events
moving through our chosen range of masses ($10^{-3}-1~\mathrm{M}_\odot$).
In particular we evaluate an expected signal of 3.9 events for $0.5~\mathrm{M}_\odot$ which,
under the hypothesis that OAB-07-N2 is a MACHO, translates into a lower limit for $f$ of about 15\%.
This outcome makes apparent the need of carrying out
similar analyses using larger sets of events, possibly across
larger fields of view, for which, besides their number,
also additional statistics may be used to disentangle the
MACHO  from the self lensing signal.
In this perspective, the observational campaign PAndromeda \citep{lee12}
promises to mark an important further step
into the understanding of this issue.

\acknowledgments
We thank M. Fitzpatrick for his support with IRAF.
AG was supported by NSF grant AST-1103471.
MD is thankful to Qatar National Research Fund (QNRF), 
member of Qatar Foundation, for support by grant NPRP 09-476-1-078.
PJ acknowledges support by the Swiss
National Science Foundation.

\bibliographystyle{apj}
\bibliography{biblio}

\end{document}